\documentclass{emulateapj}

\usepackage{amsmath}
\usepackage{graphicx}
\usepackage{epstopdf}
\usepackage{mathrsfs}

\shorttitle{Transient chaos in planetary feeding zones}
\shortauthors{Kov\'acs \& Reg\'aly}

\begin{document}

\title{Transient chaos and fractal structures in planetary feeding zones}

\author{T. Kov\'acs\altaffilmark{1} and Zs. Reg\'aly}
\affil{Konkoly Observatory of the Hungarian Academy of Sciences, H-1525 Budapest P.O. Box 67, Hungary}

\altaffiltext{1}{University of Applied Sciences, H-1148 Budapest, Nagy Lajos kir. \'utja 1-9., Hungary}

\begin{abstract}
The circular restricted three body problem is investigated in the context of accretion and scattering processes. In our model a large number of identical non-interacting mass-less planetesimals are considered in planar case orbiting a star-planet system. This description allows us to investigate in dynamical systems approach the gravitational scattering and possible captures of the particles by the forming planetary embryo. Although the problem serves a large variety of complex motion, the results can be easily interpreted because of the low dimensionality of the phase space. We show that initial conditions define isolated regions of the disk, where accretion or escape of the planetesimals occur, these have, in fact, a fractal structure. The fractal geometry of these ''basins'' implies that the dynamics is very complex. Based on the calculated escape rates and escape times, it is also demonstrated that the planetary accretion rate is exponential for short times and follows a power-law for longer integration. A new numerical calculation of the maximum mass that a planet can reach (described by the expression of the isolation mass) is also derived. 
\end{abstract}


\section{Introduction}\label{sec:intro}

In recent decades many authors studied the planetary growth using either the N-body approach or the statistical 'particle-in-box' approach assuming Rayleigh distribution functions of planetesimals \citep{nak1989,nis1983,wet1985,oht2002,raf2003}. All these works deal with the accretion rate onto a larger planetary embryo and scattering among the planetesimals in the disk, in which two- or three-body dynamics is considered depending on the velocity dispersions of the planetesimals \citep{wet1989,gre1990,gre1978,wet1989,ste2000,oht2002,orm2010}. 

For this investigation setting the origin of the reference frame to the planet and scaling the equations with $\mu^{1/3},$ where $\mu\ll 1$ is the planet-to-mass ratio. These make the Hill's approximation an ideal model to study the interaction between the planet and a particle in the circular restricted three body problem  (CRTBP) \citep{mur1999}. However, in this context the dynamics can be investigated only locally. In this approximation no azimuthal dependence of the dynamics appears. The initial conditions that lead to different end states of the particles far from the embryo's position, remain hidden in the whole configuration space. In order to map the x-y plane around the central star one needs to integrate the equations of motion of the CRTBP.

To estimate the maximum mass of a forming planet, a simple geometrical assumption has been used so far \citep{lis1993,arm2010}. The classical picture states that the whole mass that a growing planet can collect is the area of an annulus around the planet's orbit, the so-called \textit{feeding zone,} multiplied by the surface density of the planetesimals corrected by a numerical factor. However, it is known that the accretion process is very complex \citep{nis1983,pet1986,spa1994,cor1999,waa2005,orm2010} presenting chaotic zones in the feeding zone. Furthermore, there are particles in the feeding zone excluded from the accretion process, they never reach the embryo's surface but remain bounded for very long integration times.

In this Letter the CRTBP has been investigated in the context of planetesimals' accretion and escape, focusing on short- and long-time dynamics. We found that the initial conditions describing different destinations of the planetesimals in the feeding zone form fractal structure. This complex geometry is strongly related to the chaotic dynamics and, therefore, the classical formula of the maximum reachable mass should be corrected according to this structure. We also persent a general method, for a given mass parameter ($\mu=0.001$), which demonstrates how the fractal geometry can be taken into account in order to estimate the amount of the accreted planetesimals.

\section{Escapes from dynamical systems}\label{sec:escape}

\begin{figure*}
\centering
\includegraphics[width=0.97\textwidth]{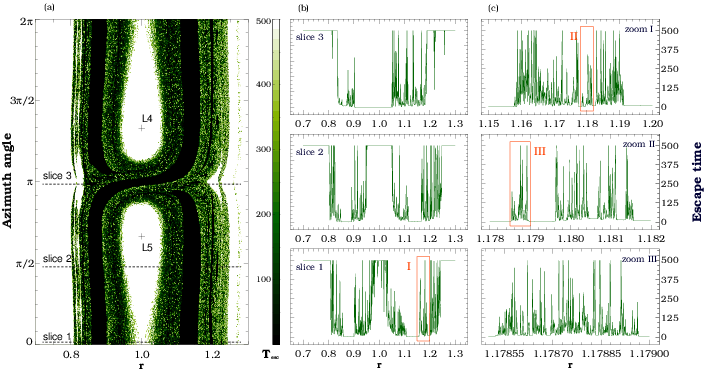}
\caption{Escape times indicated with different colors from the annulus $0.7\leq r \leq 1.3.$ The integration time is $T=500\,\text{orbital revolution}$ of the planet. Filamentary structure is formed at the border of the white regular domains (high escape times), the black trails correspond to very fast escape (accretion). Panel (b) shows the spike structure of the escape times in radial direction obtained at different azimuthal positions in panel (a). The number of long lived trajectories increases when approaching the edge of regular plateaus. Panel (c) contains magnifications of slice 1 denoted by red rectangles (I,II,III).\label{fig:esctimes}}
\end{figure*}

There are two major groups of conservative dynamical systems, as is the CRTBP, where escapes are possible. The first one is called an \textit{open system,} in which, for certain energy levels, the equipotential surfaces open up and the originally bounded domain of the phase space extends to infinity. Such open systems are frequent in dynamical astronomy \citep{ben1998,ern2008,ast2003,sim2000,agu2001,har2011,con2002}. The escape process from the CRTBP proceeds in two stages: (i) escape through openings in the equipotential surfaces around the saddle points of the potential (collinear Lagrangian points) and (ii) two-body encounters that induce scattering of the planetesimals into the escaping part of the phase space, referred to as the chaotic scattering process \citep{eck1986,jun1987,ble1989,tel2006,ott2002,pet1986}.

The second group of dynamical systems that show escaping behavior contains the so-called \textit{leaky chaotic systems}. In this concept an originally closed system is leaked with an arbitrary (but small) hole somewhere in phase (or configuration) space and the probability of escaping trajectories can be measured through that hole \citep{alt2013}. This problem is "leaky" in that planetesimals can collide with the planet embryo, and hence that the planet's physical extent represents a hole in the configuration (or phase) space through which trajectories can escape the system. Therefore, there is a second escape route from the system (accretion) that is not present in the system where the planet and star are point particles.

These two groups of the escaping dynamics are related via the theory of transient chaos \citep{lai2011}. If there are more than one leak in the system, they share all the dynamical properties of finite-time chaotic behavior, i.e. the average lifetime of chaos, the escape rate, the chaotic saddle, average Lyapunov exponents, etc. \citep{alt2013}. Interestingly, these results hold in our case as well, i.e. if the system contains an escape route to infinity and also a leak, the dynamics behaves as if it would be a two-leak-system. Finite-time chaotic behavior in dynamical astronomy may appear in scattering processes such as, close encounters, temporary captures \citep{iwa2007,lee2007}, moon-ring \citep{jac2012} and planet-planetesimal interactions. 

Transient chaos serves two types of escape processes: short- and long-time escapes. Quantitatively, the exponential decay of the survival probability for short-time escapes can be described as \citep{alt2013}:
\begin{equation}
P(t)\sim e^{-\kappa t},
\label{eq:exp}
\end{equation} 
where the escape rate $\kappa$ is an invariant quantity for a large number of initial conditions describing the short-time escape dynamics.

For long-time escapes the decay of trajectories becomes power-law instead of the exponential decay equation~(\ref{eq:exp})
\begin{equation}
P(t)\sim t^{-\sigma},
\label{eq:pow} 
\end{equation}
where $\sigma$ denotes an algebraic decay exponent of chaotic trajectories that might come close to regular domains and mimic ordered motion for very long times.

\section{Numerical investigation of escaping dynamics}\label{sec:numerics}

To study the escaping dynamics in CRTBP we performed numerical calculations of a large number ($3-4\times10^{5}$) of non-interacting mass-less planetesimals placed initially on a ring ($r\in[0.7,1.3],\phi\in[0,2\pi]$) around the barycenter of the system. Their initial conditions define circular Keplerian orbits. We consider co-rotating frame with the Keplerian angular velocity of the planet. The length unit equals to the planet-star distance, the time unit is the orbital period, and the sum of the masses equals to 1. In these units the gravitational constant is 1, and the dynamics of planetesimals is governed by the dimensionless equations of motion \citep{szeb1967}
\begin{equation}
\begin{split}
\frac{\mathrm{d}r}{\mathrm{d}t}&=v,\quad\frac{\mathrm{d}v}{\mathrm{d}t}=\frac{L^{2}}{r^{3}}-\frac{\partial \Phi}{\partial r},\\
r\frac{\mathrm{d}\phi}{\mathrm{d}t}&=L,\quad\frac{\mathrm{d}L}{\mathrm{d}t}=-\frac{\partial \Phi}{\partial \phi},
\end{split}
\label{eq:motion}
\end{equation}
where $r,$ $v,$ and $L$ are the barycentric distance, radial velocity, and angular momentum, respectively. $\Phi$ denotes the stellar and planetary gravitational potential.

The escape of a planetesimal from the system can be interpreted as the accretion onto the planetary surface or scattering out of the system by the planet. In our model the above processes occur when
\begin{itemize}
\item (accretion) a planetesimal approaches the planet closer than $0.5r_{H}$ ($r_{H}$ is the Hill radius of the planet) and its velocity is smaller than the escaping velocity at that distance. The planetesimal is removed from the system and its initial position is marked as a point of the leak formed by the planet in the initial annulus;
\item (scattering) the planetesimal is perturbed to a hyperbolic orbit caused by a close encounter with the planet. In this case the planetesimal is also removed from the system and its initial position is added to the escape channel to infinity. 
\end{itemize}

Since almost all planetesimals starting from the feeding zone leave the system (scattered out or accreted) within couple of hundreds orbit in agreement with \citet{par2007}, calculations were performed up to 500 orbital revolutions of the planet for short time dynamics and 10000 orbits for long-time dynamics. The integrator is a fourth-order adaptive step size Runge-Kutta scheme with a maximum relative energy error of $10^{-8}.$ We did not use the regularized equations of motion because close encounters due to the accretion conditions do not appear in this problem. We considered no mass growth of the embryo in time as it accretes particles. In order to save computation time, planetesimals on hyperbolic orbits were followed until their radial distance exceeded $r=5$ \citep{nag2005}. Some test runs show that there are planetesimals with much longer life-times, and these and the mechanism of the dynamics behind this phenomenon will be discussed in the next section. In the numerical simulations the mass ratio is $\mu=0.001,$ which is close to the parameter of the Sun-Jupiter system. 
\section{Results}\label{sec:res}

\begin{figure*}
\centering
\includegraphics[width=0.97\textwidth]{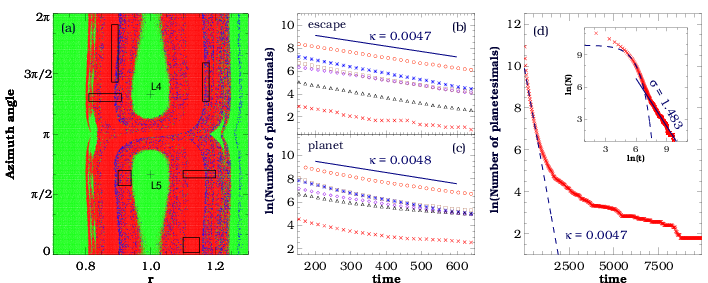}
\caption{''Basins of attraction'' and escape rates. (a) Different colors represent different final states of planetesimals starting from the initial conditions $(r,\theta).$ Green: bounded motion, red: accretion onto the planet, and blue: escape from the system. The integration time is $T=500\,\text{revs.}$ The filamentary structure is similar to that in Fig.~\ref{fig:esctimes}. (b-c) Short-time escape dynamics for planetesimals starting from rectangles in panel (a). Six different ensembles of planetesimals show the same escape rate, $\kappa=0.0047,$ independent of that how they leave the system (accreted or escaped to infinity) (d). Simulations for longer integration time ($T=10000\,\text{rev}$) show power-law decay of escaping planetesimals. This phenomenon is related to the trajectories that may come close to the border of regular domains. The decay rate is described by the exponent $\sigma=1.483.$ \label{fig:boa}}
\end{figure*}

In what follows, we present our numerical results on the short- and long-time escape dynamics of planetesimals and their relation to the geometrical structure of the disk.

Figure \ref{fig:esctimes}(a) shows the contour plot of escape times from a given initial position in the planetesimal disk. Next to this panel the color-bar indicates the color codes for different escape times (up to $t=500\,\text{rev}$). Monitoring the escape times means tracing of how long an individual planetesimal lives as a member of the disk before leaving it either to infinity or accreted to the planet's surface.

As can be seen in panel (a), two sharp zones (black) are formed around the planet's orbit corresponding to very low escape times, $\lesssim 50\text{ revolutions}.$ 

The most interesting domain is, however, the edge of two low-escape time trails at the inner (close to the \textit{horseshoe region}\footnote{Part of the disk close to the planetary orbit, where the planetesimals' motion encompasses the Lagrangian points $L_{3},\,L_{4},\,L_{5}.$ These planetesimals cannot enter the Hill sphere of the planet.}) and at the outer part of the feeding zone (where Keplerian dynamics is not affected by the forming planet). One may think that initial positions with the same order of escape times are distributed randomly or at least ''scattered'' in the annulus. Our simulation demonstrates that on the contrary, the escape times show a filamentary fractal structure at the borders of the regular domains.

To check the fine structure of panel (a) we cut the disk in radial direction at three different azimuthal positions, $\theta=0.05,\,1.5,\,3.05$ radians (dashed lines). Escape times for these slices are plotted in Figure~\ref{fig:esctimes}, panel (b). One can observe the spiky structure of the plot. Plateaus represent the non-escaping domains. Note that the closer the initial position to the edge of the bounded region (where the planetesimals are neither accreted nor got escaped), the longer the escape time. Other interesting feature can be obtained when the resolution is refined at a given region, panel (c). We can conclude that the structure is self-similar, i.e. between spiky parts one can always find shorter or longer segments corresponding to very low escape times. This self-similarity is the fingerprint of chaotic dynamics while close encounters between the planet and planetesimals take place.

A more informative picture can be drawn if we separate the initial conditions depending on the final state of the motion. Figure \ref{fig:boa}(a) depicts the initial conditions of bounded (green), escaped (blue), and accreted (red) planetesimals with different colors. We can consider these subsets of the disk as the ''basins of attraction'', say, the red dots form the basin of attraction of the planet, i.e. all these planetesimals hit the planet's surface sooner or later. Blue dots mark the basin of attraction of the infinity. The planet's feeding zone can, therefore, be thought as a subset of the Keplerian disk between $0.8\lesssim r \lesssim 1.23,$ excluding the horseshoe region and trajectories that escape the system.

The similarity with Fig.~\ref{fig:esctimes} is evident, some of the escaped and bounded trajectories seem to overlap the filamentary structure of high escape times. The picture is, however, more complex. There are also initial conditions from the red domain (Fig.~\ref{fig:boa}(a)) that survive long integration time before they are accreted as we will see later. 

Figure~\ref{fig:boa}(b) and (c) contain the number of surviving planetesimals vs. time of the six black rectangles in Fig.~\ref{fig:boa}(a). The log-linear plot indicates that for shorter times the escape is exponential as is expected from the dynamical system theory. The average life-time, $\tau,$ that a planetesimal spends in the system, can be approximated by the reciprocal of the escape rate $\kappa.$ Separating the scattered (panel (b)) and the accreted (panel (c)) trajectories one may notice the same slope of the exponential decays. This is the consequence of a fundamental phenomenon in leaky chaotic systems. Namely, the escape rate is an invariant quantity but the amount of planetesimals outflowing through the different leaks, in our case the planet and the gate to the infinity, might be different \citep{alt2013}. For example, in case of box $r\in[1.1,1.2],\,\theta\in[2,2.2]$ 48620 planetesimals hit the planet and 1379 planetesimals leave the system with the same rate (and 1 remains bounded), $\kappa\approx 0.0047,\,\tau\approx 200\text{ revs},$ (see equation~(\ref{eq:exp})).

One can also identify the power-law decay for longer integration times, see Fig.~\ref{fig:boa}(d). The tail of the survival probability always represents the sticky orbits that come close to the regular domains of the phase space. The decay rate $\sigma$ defined by equation~(\ref{eq:pow}) is found to be 1.483.

\begin{figure}
\centering
\includegraphics[width=0.48\textwidth]{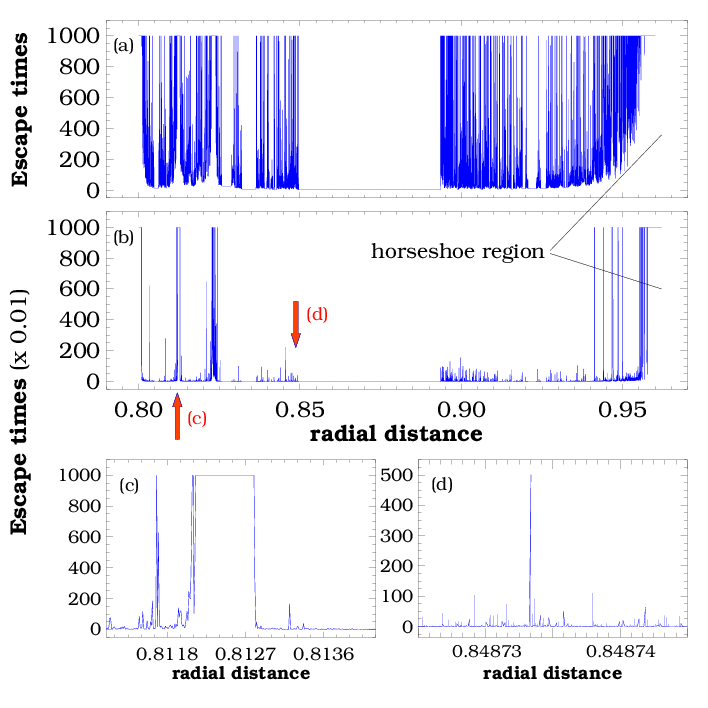}
\caption{Escape times for different time scales, T=$10^3$ (a) and T=$10^5$ (b) revolutions. For the relevant time scales of the system, i.e. until accretion takes place, fractality exists. The plateau represents a high-order mean motion resonance (c) and a lonely spike appears (d) close to the stable manifold of an unstable periodic orbit.}\label{fig:longt}
\end{figure}

Figure~\ref{fig:longt} compares the spiky structure of slice 2 (see Fig.~\ref{fig:esctimes}) for different integration times. Short plateaus still appear for $10^6$ revolutions, panels (b,c). This feature is responsible for higher order mean motion resonances where planetesimals are trapped and their motion remains bounded. At some radial distances, relatively high lonely spikes are also visible in the middle of triangle-like shape formed by shorter escape times, panel (d). This phenomenon is related to unstable periodic orbits and the high escape time indicates that we started the planetesimal close to the stable manifold of the hyperbolic periodic orbit (HPO). The union of these HPOs is the backbone of the chaotic saddle and controls the transient dynamics in the system. Note that this behavior was also found by \citet{pet1986} (Figure 6) [called \textit{transition zones}] but they did not recognize it as a basis of transient chaos.

With this procedure we can estimate numerically the ratio of the accreted and scattered planetesimals in the planetary feeding zone. It can be shown that the filamentary structure, which is actually a fractal set, is an invariant of the dynamics, i.e. the fractal dimension of the basins of attraction is constant \citep{lai2011}. This results in the extended formula of the isolation mass -- the mass of all the planetesimals that are accreted to the planet, $M_{\mathrm{iso}}$ 
\begin{equation}
M_{\mathrm{iso}}=4\pi a C r_{H} \Sigma_{p},
\label{eq:m_iso}
\end{equation}
where $C r_{H}=\Delta a$ is the half-width of the feeding zone, $a$ is the orbital radius, and $\Sigma_{p}$ denotes the surface density of planetesimals. $C$ has different values in the literature, e.g. 2.5 \citep{pet1986}, $2\sqrt{3}$ \citep{lis1993}, 2.3 \citep{arm2010}. In equation~(\ref{eq:m_iso}) the factor $C$ describes a regular annulus shape without planetesimals scattered from the systems or trapped in higher order mean motion resonances.

\begin{figure}
\centering
\includegraphics[width=0.45\textwidth]{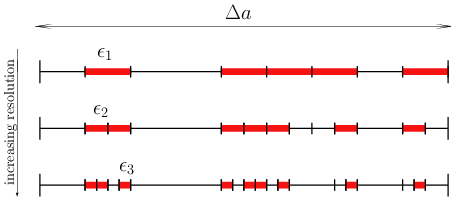}
\caption{Schematic picture on the number of segments covering the half-width ($\Delta a$) of the feeding zone. The finer the resolution, the more the segments cover the filamentary structure. The power-law link between $N$ and $\epsilon$ gives the fractal dimension of the filamentary structure.}\label{fig:res}
\end{figure}

Therefore, the geometrical factor, $C,$ must be revisited in equation~(\ref{eq:m_iso}). Let us suppose we can measure the ''space filling factor'' of the filamentary structure for the accreted planetesimals, i.e. the fractal dimension is known. Basically, we want to know the number $N,$ of $\epsilon$-long line segments that cover the accretion part of $\Delta a,$ see the schematic picture in Fig.~\ref{fig:res}. This procedure is known as \textit{box counting dimension.} Let $\epsilon/\Delta a$ be the dimensionless length of a line segment, where $\epsilon$ varies through several magnitudes, and, therefore, can be thought as of the measure of resolution. The number of line segments, $N,$ depending on $\epsilon$ reads
\begin{equation}
N\bigg(\frac{\epsilon}{\Delta a}\bigg)\sim \bigg(\frac{\epsilon}{\Delta a}\bigg)^{-D},
\label{eq:n_eps}
\end{equation}
where $D$ denotes the fractal dimension.\footnote{More precisely, $D$ is the information dimension but it can be approximated by the fractal dimension.\citep{tel2006}} From equation~(\ref{eq:n_eps}) the whole length covered by accreted planetesimals in radial direction can be obtained simply 
\begin{equation}
L\sim\epsilon N=\epsilon\bigg(\frac{\epsilon}{\Delta a}\bigg)^{-D}=\epsilon^{1-D}(\Delta a)^{D}.
\label{eq:length}
\end{equation}
Following equation~(\ref{eq:length}) the half-width $\Delta a$ in equation~(\ref{eq:m_iso}) should be changed to $\epsilon^{1-D}(\Delta a)^{D}$, and, consequently, $C=\Delta a/r_{H}$ becomes
\begin{equation}
C=\frac{(\Delta a)^{D}\epsilon^{1-D}}{r_{H}}=\frac{\Delta a}{r_{H}}(\Delta a)^{D-1}\epsilon^{1-D},
\label{eq:c_gen}
\end{equation}
where $\epsilon / \Delta a < 1,\,0\leq 1-D < 1,$ and we can write a new, generalized multiplication factor, $C_{\mathrm{gen}},$ as
\begin{equation}
C_{\mathrm{gen}}=C\bigg(\frac{\epsilon}{\Delta a}\bigg)^{1-D},\quad \text{where}\quad D=1-\frac{\kappa}{\overline{\lambda}},
\label{eq:kantz}
\end{equation}
where $D$ is the fractal dimension in radial direction, $\kappa$ denotes the escape rate, and $\overline{\lambda}$ the average Lyapunov exponent (LE) of the transient trajectories. Thus, we have a new numerically determined geometrical factor in the formula of the isolation mass which is related to the dynamics of the system through the LE. The link between $D,\;\kappa,\;\text{and}\;\lambda$ is the Kantz-Grassberger relation \citep{kan1985}. 

In equation~(\ref{eq:kantz}) the original $C$ is an upper limit to the general multiplication factor $C_{\mathrm{gen}}.$ Additionally, one can clearly see that $C_{\mathrm{gen}}$ tends to $C$ when $\kappa\to 0.$  It means when the planetary perturbation is small ($\mu\to 0$), i.e. in the limiting case when no planet is present, the problem reduces to a simple two-body problem [star-and-planetesimal]. Consequently, all the planetesimals revolve on Keplerian orbit defined by initial conditions and no scattering takes place ($\kappa=0$), hence $C_{\mathrm{gen}}=C$. A detailed analysis of the $\mu$ dependence of $C$ through $D$ will be presented in a forthcoming paper.

In practice, first the half-width of the feeding zone ($\Delta a$) must be computed. It is a simple procedure where one needs to find the inner and outer border of that part of the disk where the main bounded plateaus start. The Hill's radius is defined by the mass parameter ($r_{H}\sim\mu^{1/3}$) of the system. Next step is the fractal dimension $D$ which can be obtained from the covering $\epsilon$-line segments. Having $\Delta a, r_{H}, \epsilon,\text{ and } D$ finally we can compute $C_{\mathrm{gen}}$ from equation~(\ref{eq:kantz}). $C_{\mathrm{gen}}$ for $\mu=0.001$ is found to be 1.693.

\section{Summary and conclusion}

In this Letter we presented a dynamical investigation of an ensemble of non-interacting planetesimals in circular restricted-three body problem for the mass parameter $\mu=0.001.$ 

Numerical calculations show that the initial conditions of accreted and scattered planetesimals in the planetary feeding zone form filamentary structure which is a fractal set along the radial direction. This structure indicates the complex motion of planetesimals having close encounter with the planet. We can also identify self-similarity in escape time distribution along the radial direction for relevant time scales of the system. This implies that transient chaos is responsible for the short- and long-time escape dynamics. 

It is possible to separate the initial conditions depending on their final destinations. Therefore, we proposed to identify the fractal dimension along the radial direction of the ''basins of attraction'' and introduced a simple correction with this measure in the formula of the isolation mass. Nevertheless, the fractal dimension is in close relation with dynamical parameters of the systems, the escape rate and the average Lyapunov exponent. In this context, we can say that the feeding zone and the isolation mass are strongly related through the chaotic dynamics of the system. We also demonstrated that for low-mass planets  the new formula tends to the classical expression.

\acknowledgments
We are grateful to T. T\'el and E.G. Altmann for the fruitful discussions and L.L. Kiss to clarify the text. This work has been supported by the Lend\"ulet-2009 Young Researchers' Program of the HAS, the ESA PECS Contract No.~4000110889/14/NL/NDe, the Hungarian OTKA Grant No.~K101393, and the Momentum grant of the MTA CSFK Lend\"ulet Disk Research Group.

\clearpage

\begin{thebibliography}{}
\bibitem[Aguirre et al.(2001)]{agu2001} Aguirre, J., Vallejo, J. C., Sanju\'an, M. A. F. 2001, \pre, 64, 066208
\bibitem[Altmann et al. (2013)]{alt2013} Altmann, E. G., Portela, J. S. E., T\'el, T. 2013, RvMP, 85, 869
\bibitem[Armitage(2010)]{arm2010} Armitage, P. J. 2010, Astrophysics of Planet Formation, Cambridge University Press 
\bibitem[Astakhov et al.(2003)]{ast2003} Astakhov, S., Burbanks, A., Wiggins, S., Farrelly, D. 2003, \nat, 423, 264
\bibitem[Benet(1998)]{ben1998} Benet, L., Seligman, T. H., Trautmann, D. 1998, Cel. Mech and Dyn. Astron., 71, 167
\bibitem[Bleher et al.(1989)]{ble1989} Bleher, S., Ott, E., Grebogi, C. 1989, \prl, 63, 919
\bibitem[Contopoulos(2002)]{con2002} Contopoulos, G. 2002, Order and Chaos in Dynamical Astronomy, Springer Science \& Business Media
\bibitem[Cordeiro et al.(1999)]{cor1999} Cordeiro, R. R.; Martins, R. Vieira; Leonel, E. D. 1999, \aj, 117, 1634
\bibitem[Eckhardt \& Jung(1986)]{eck1986} Eckhardt, B., Jung, C. 1986, JPhA, 19, 829
\bibitem[Ernst et al.(2008)]{ern2008} Ernst, A., Just, A., Spurzem, R., Porth, O. 2008, \mnras, 383, 897
\bibitem[Goldreich et al.(2004)]{gol2004} Goldreich, P., Lithwick, Y., Sari, R. 2004, \araa, 42, 549
\bibitem[Greenberg(1978)]{gre1978} Greenberg, R. 1978, Icarus, 33, 62
\bibitem[Greenzweig \& Lissauer(1990)]{gre1990} Greenzweig, Y., Lissauer, J. J. 1990, Icarus, 87, 40
\bibitem[Harsoula \& Contopoulos(2011)]{har2011} Harsoula, M., Kalapotharakos, C., Contopoulos, G. 2011, \mnras, 411, 1111
\bibitem[Ida \& Nakazawa(1989)]{ida1989} Ida, S., Nakazawa, K. 1989, \aap, 224, 303
\bibitem[Iwasaki \& Ohtsuki (2007)]{iwa2007} Iwasaki, K., Ohtsuki, K. 2007, \mnras, 377, 1763
\bibitem[Jackson \& Wyatt (2012)]{jac2012} Jackson, A.P., Wyatt, M.C. 2012, \mnras, 425, 657
\bibitem[Jung(1987)]{jun1987} Jung, C. 1987, JPhA, 20, 1719
\bibitem[Kantz \& Grassberger(1985)]{kan1985} Kantz, H., Grassberger, P. 1985, Physica D, 17, 75
\bibitem[Kary et al.(1993)]{kar1993} Kary, D. M., Lissauer, J. J., Greenzweig, Y. 1993, Icarus,
\bibitem[Lai \& T\'el(2011)]{lai2011} Lai, Y.-C., T\'el, T. 2011, Transient Chaos, Springer Science \& Business Media
\bibitem[Lee et al. (2007)]{lee2007} Lee, E.A., Astakhov, A., Farrelly, D. 2007, \mnras, 379, 229
\bibitem[Lissauer(1993)]{lis1993} Lissauer, J. J. 1993, \araa, 31, 129
\bibitem[Lissauer \& Stewart(1993)]{lisste1993} Lissauer, J. J., Stewart, G. R. 1993, APS Conf., 36, 217
\bibitem[Murray \& Dermott(1999)]{mur1999} Murray, C., Dermott, S. F. 1999, Solar System Dynamics, Cambridge University Press 
\bibitem[Nakazawa et al.(1989)]{nak1989} Nakazawa, K., Ida, S., Nakagawa, Y. 1989, \aap, 220, 293
\bibitem[Nagler (2005)]{nag2005} {Nagler, J.} 2005, PhRvE, 71, 6227
\bibitem[Nishida(1983)]{nis1983} Nishida, S. 1983, Progress of Theoretical Physics, 70, 93
\bibitem[Ohtsuki et al.(2002)]{oht2002} Ohtsuki, K., Stewart, G. R., Ida, S. 2002, Icarus, 155, 436
\bibitem[Ormel \& Klar(2010)]{orm2010} Ormel, C. W., Klahr, H. H. 2010, \aap, 520, 43
\bibitem[Ott(2002)]{ott2002} Ott, E. 2002, Chaos in Dynamical Systems, Cambridge University Press
\bibitem[Paardekooper(2007)]{par2007} Paardekooper, S.-J. 2007, \aap, 462, 355
\bibitem[Petit \& H\'enon(1986)]{pet1986} Petit, J.-M., Henon, M. 1986, Icarus, 66, 536
\bibitem[Rafikov(2003)]{raf2003} Rafikov, R. R. 2003, \apj, 125, 922
\bibitem[Sim\'o \& (2000)]{sim2000} Sim\'o, C., Stuchi, T. J. 2000 Phyisica D, 140, 1
\bibitem[Spahn et al. (1994)]{spa1994} Spahn, F., Scholl, H., Hertzsch, J-M. 1994, Icarus, 111, 514
\bibitem[Szebehely(1967)]{szeb1967} Szebehely, V. 1967, Theory of orbits, Academic Press
\bibitem[Stewart \& Ida(2000)]{ste2000} Stewart, G. R., Ida, S. 2000, Icarus 143, 28
\bibitem[T\'el \& Gruiz(2006)]{tel2006} T\'el T., Gruiz, M. 2006, Chaotic Dynamics, Cambridge University Press 
\bibitem[Waalkens et al. (2005)]{waa2005} Waalkens, H.; Burbanks, A.; Wiggins, S. 2005, \mnras, 361, 763
\bibitem[Weidenschilling \& Davis(1985)]{wei1985} Weidenschilling, S. J., Davis, D. R. 1985, Icarus, 62, 16
\bibitem[Wetherill \& Cox(1985)]{wet1985} Wetherill, G. W., Cox, L. P. 1985, Icarus, 63, 290
\bibitem[Wetherill \& Stewart(1989)]{wet1989} Wetherill, G. W., Stewart, G. R. 1989, Icarus, 77, 330
\end{thebibliography}
\end{document}